\documentclass[pra,twocolumn]{revtex4}

\usepackage{graphicx}
\usepackage{array}
\usepackage{color}
\usepackage{amssymb}
\usepackage{amsmath}

\begin{document}

\title{Exact solution of the 1D time-dependent Schr\"odinger equation for the emission of quasi-free electrons from a flat metal surface by a laser}

\author{Ovidiu Costin}
\author{Rodica Costin}
\affiliation{Ohio State University,Columbus, OH 43210, USA}

\author{Ian Jauslin}
\affiliation{Princeton University, Princeton, NJ 08544, USA}

\author{Joel L. Lebowitz}
\affiliation{Rutgers University, Piscataway, NJ 08854, USA}

\begin{abstract}
We solve exactly the one-dimensional Schr\"odinger equation for $\psi(x,t)$ describing the emission of electrons from a flat metal surface, located at $x=0$, by a periodic electric field $E\cos(\omega t)$ at $x>0$, turned on at $t=0$.
We prove that for all physical initial conditions $\psi(x,0)$, the solution $\psi(x,t)$ exists, and converges for long times, at a rate $t^{-\frac32}$, to a periodic solution considered by Faisal et al. (Phys. Rev. A {\bf 72}, 023412 (2005)).
Using the exact solution, we compute $\psi(x,t)$, for $t>0$, via an exponentially convergent algorithm, taking as an initial condition a generalized eigenfunction representing a stationary state for $E=0$.
We find, among other things, that: (i) the time it takes the current to reach its asymptotic state may be large compared to the period of the laser;
(ii) the current averaged over a period increases dramatically as $\hbar\omega$ becomes larger than the work function of the metal plus the ponderomotive energy in the field. For weak fields, the latter is negligible, and the transition is at the same frequency as in the Einstein photoelectric effect;
(iii) the current at the interface exhibits a complex oscillatory behavior, with the number of oscillations in one period increasing with the laser intensity and period.
These oscillations get damped strongly as $x$ increases.
\end{abstract}

\maketitle

\section{Introduction}
\indent There have been many advances in recent years in the development and application of short intense laser pulses to produce femto-second and even atto-second beams of electrons from metallic surfaces \cite{HKK06,SKH10,BGe10,KSH11,KSe12,THH12,HSe12,PPe12,HBe12,PSe14,HWR14,EHe15,BBe15,YSe16,FSe16,RLe16,FPe16,LJ16,HKe17,SSe17,PHe17,Je17,WKe17,KLe18,LCe18,SMe18}.
A full microscopic description of the short-time behavior of the emission process is therefore highly desirable.

\indent In this note, we present, for the first time, an exact solution for the time-dependent Schr\"odinger equation describing the emission of electrons from a flat metal surface by an oscillating electric field.
We use the Sommerfeld model of quasi-free electrons with a Fermi distribution of energies, confined by a step potential $U=\mathcal E_F+W$, where $\mathcal E_F$ is the Fermi energy and $W$ is the work function of the metal.
This setup was first used by Fowler and Nordheim \cite{FN28} in 1928 for a time-independent field, and is commonly used as a model for the process of emission, both for a constant and an oscillating field \cite{FN28,FKS05,HKK06,KSH11,YGR11,KSe12b,PA12,YHe13,CPe14,ZL16,Fo16,Je17,KLe18}.
In both cases one imagines the metal occupies the half space $x<0$, and focuses attention on electrons, part of the Fermi sea, moving from the left towards the metal surface at $x=0$.
These are described by a wave function $e^{ikx}$, $k>0$, $x<0$ and have energy $\frac12k^2$ (in atomic units).
In the sequel, we shall generally consider values of $k$ such that $\frac12k^2=\mathcal E_F$.
The field is described classically.
\medskip

\indent The time evolution of the wave function of an electron in such a beam subjected to an oscillating field for $x\geqslant 0$, is described by the one dimensional Schr\"odinger equation: for $x\in\mathbb R$ and $t>0$,
\begin{equation}
  i\partial_t\psi(x,t)=-\frac12\Delta\psi(x,t)+\Theta(x)(U-Ex\cos(\omega t))\psi(x,t)
  \label{schrodinger}
\end{equation}
where $\Theta(x)=0$ for $x<0$ and $\Theta(x)=1$ for $x>0$, $E$ is the electric field perpendicular to the surface, $\frac\omega{2\pi}$ is the frequency and we are using atomic units $\hbar=m=1$.
We note that in experiments one usually applies the laser field to a sharp tip in order to enhance the strength of the field.
One also includes a carrier wave envelope.
Here we ignore these as well as the Shottky effect.
Including them would greatly complicate the problem.
We believe that the simpler model considered here already captures many of the relevant physical phenomena so we focus on its exact solution.
The values of the field we use in our computations are those generally used for the enhanced field at a sharp tip.
The short time behavior would be the same as if the field was cut off after some time $t_0$.

\indent In the absence of an external field, $E=0$, the Schr\"odinger equation (\ref{schrodinger}) has a ``stationary" solution $e^{-i\frac12k^2t}\varphi_0(x)$ in which there is, for $k^2<2U$, a reflected beam of the same energy and intensity as the incoming beam $e^{ikx}$ and an evanescent, exponentially decaying tail on the right.
The requirement of continuity of $\psi$ and its spatial derivative at $x=0$ then gives \cite{Je17}
\begin{equation} \label{initial}
  \varphi_0(x)=
  \left\{\begin{array}{>\displaystyle ll}
    e^{ikx}+\frac{ik+\sqrt{2U-k^2}}{ik-\sqrt{2U-k^2}}e^{-ikx}&\mathrm{for\ }x<0
    \\[0.5cm]
    \frac{2ik}{ik-\sqrt{2U-k^2}}e^{-\sqrt{2U-k^2} x}&\mathrm{for\ }x>0
  \end{array}\right.
\end{equation}
The current,
\begin{equation}
  j(x,t):=\mathcal Im(\psi^*(x,t)\partial_x\psi(x,t))
  \label{current}
\end{equation}
is zero and no electrons leave the metal.
\medskip

\indent In \cite{CCe18} we solved the time-dependent Schr\"odinger equation (\ref{schrodinger}) for any initial condition, including $\varphi_0(x)$, for the constant field.
This corresponds to setting $\omega=0$ in (\ref{schrodinger}).
We showed that $\psi(x,t)$ converges, as $t\to\infty$, to the well known Fowler-Nordheim (FN) solution \cite{FN28} for emission by a constant field.
FN assumed a solution of (\ref{schrodinger}), with $\omega=0$, of the form $e^{-\frac12k^2t}\varphi_E(x)$, so that $\varphi_E(x)$ satisfies the equation
\begin{equation}
  -\frac12\Delta\varphi_E+\Theta(x)(U-Ex)\varphi_E=\frac12k^2\varphi_E
  .
\end{equation}
The solution $\varphi_E(x)$ has the form $\varphi_E(x)=e^{ikx}+R_Ee^{-ikx}$ for $x<0$ and an Airy function expression for $x>0$.
The FN computation of the tunneling current from $\varphi_E(x)$ via (\ref{current}), is still the basic ingredient for the analysis of experiments at present \cite{Je17}.
The main modification is the use of the Schottky factor \cite{Je17,Fo16}, rounding off the barrier at $x=0$, which, as already noted by FN, is only important for $\frac12k^2\sim U$.

The rate of convergence of the solution of the time-dependent Schr\"odinger equation (\ref{schrodinger}), with $\omega=0$ and an initial condition of the form (\ref{initial}), to the FN solution was shown in \cite{CCe18} to be like $t^{-\frac32}$.
Surprisingly, the deviation of the current from the FN solution becomes quickly very small, so the ``effective'' time of approach to the FN solution was found to be, for realistic values of the parameters, of the order of femtoseconds.
It is therefore not significant for emission in constant fields acting over much longer times.

\indent Here, we investigate solutions of (\ref{schrodinger}) with $E>0$, $\omega>0$ and general $\psi(x,0)$.
This covers a wide range of physical situations, depending on $\omega$ and $E$, ranging from mechanically produced oscillating fields to those produced by lasers of high frequency.
As the Keldysh parameter $\gamma:=\frac{2\omega}E\sqrt W$ increases, the process goes from tunneling to multi-photon emission \cite{EHe16}.
In situations in which the laser is turned on for only a short time, with pulses as short as a few laser periods, knowing the early time behavior is important.
In fact, we shall see later that the asymptotic approach to the periodic state considered in Faisal et al. \cite{FKS05}, which we discuss in detail in Section \ref{longtime}, can be much longer than a laser period.
Solving (\ref{schrodinger}) for $\omega\neq0$ turns out to be much more difficult than the constant field case, since here, the solution cannot be written in terms of known functions.
The very existence of physical solutions, which are bounded at infinity, is not mathematically obvious.
\medskip

\indent In addition to the work in \cite{FKS05} there have been very many studies of the emission process using various approximations \cite{Wo35,Ke45,CFe87,Tr93,HKK06,KSH11,YGR11,KSe12b,PA12,YHe13,CPe14,ZL16,KLe18}.
Of particular note is the work of Yalunin, Gulde and Ropers \cite{YGR11} who consider the same setup as we (except for a phase difference in the field).
They use various analytic approximations for obtaining solutions of the periodic type considered in \cite{FKS05}.
They also carry out numerical solutions using the Crank-Nicolson method.
This method is discussed in section \ref{Sec3} and compared to the exact result in Figure \ref{fig:cn}.
As shown there the Crank-Nicolson method is not correct for very short times.
In particular, the current does not tend to its initial value, zero, as time tends to 0, see \cite[Figure 5]{YGR11}.
\medskip

\indent The outline of the rest of the paper is as follows.
In section \ref{sec:math}, we give a brief description of the method used to solve (\ref{schrodinger}).
In section \ref{Sec3}, we present the results for the initial state $\psi(x,0)=\varphi_0(x)$ in (\ref{initial}).
In section \ref{longtime}, we describe the asymptotic form of $\psi(x,t)$ as $t\to\infty$.
The appendix contains more information about the derivation of the solution of (\ref{schrodinger}).

\section{Solution of the Schr\"odinger equation}\label{sec:math}
\indent We solve (\ref{schrodinger}) by using the one sided Fourier transforms $\hat{\psi}_-(\xi,t)=\frac{1}{\sqrt{2\pi}}\int_{-\infty}^0{\rm e}^{-i\xi x}\psi(x,t)\, dx$ and $\hat{\psi}_+(\xi,t)=\frac{1}{\sqrt{2\pi}}\int_0^{\infty}{\rm e}^{-i\xi x}\psi(x,t)\, dx$. These satisfy the equations 
\begin{equation}\label{A}
i\frac{\partial \hat{\psi}_-(\xi,t)}{\partial t}-\frac{\xi^2}2 \hat{\psi}_  
-= \frac1{\sqrt{2\pi}} \frac{\partial\psi}{\partial y}(0,t)-i\xi   \frac1{\sqrt{2\pi}} \psi(0,t) 
\end{equation}
and 
\begin{multline}\label{B}
 i\frac{\partial \hat{\psi}_+(\xi,t)}{\partial t}+ i \frac E2 \cos (\omega t)\,  \frac{\partial \hat{\psi}_+}{\partial \xi} 
-\left(\frac{\xi^2}2+U\right) \hat{\psi}_+(\xi,t)    \\=
\frac1{\sqrt{2\pi}}\frac{\partial\psi}{\partial x}(0,t) +i\xi   \frac1{\sqrt{2\pi}}  \psi(0,t)
\end{multline}
Both (\ref{A}) and (\ref{B}) admit explicit solutions for initial values $\hat\psi_\pm(\xi,0)$ and specified boundary values $\psi(0,t)$ and $\partial_x\psi(0,t)$, see the Appendix. The continuity conditions for $\psi$ and $\partial_x\psi$ at $x=0$ then lead to an integral equation for $\psi(0,t)$ of the form
\begin{equation}
  \label{eq:inteq}
  \psi(0,t)=h(t)+L\psi(0,t)
\end{equation}
where $L$ is some compact  integral operator whose expression is rather involved, see the appendix, and $h(t)$ is a function of the initial condition $\psi(x,0)$. We prove the existence and uniqueness of a physical solution of (\ref{eq:inteq}) for all $t>0$, by showing that $L$ is a contraction.
Given that solution $\psi(0,t)$, we can obtain $\psi(x,t)$ for all $x$ by direct integration.
To evaluate the solution numerically for the initial condition (\ref{initial}), we expand $\psi(0,t)$ in a Chebyshev polynomial series and identify the coefficients of this expansion. The complexity of this integral operator $L$ results in complex behavior of its solutions as discussed in the sequel.
The full mathematical proof of the existence and uniqueness of the solution of (\ref{schrodinger}) will be presented separately \cite{CCe}.

%

\section{Short time behavior}\label{Sec3}
\indent We carried out the numerical solution of the integral equation for $\psi_0(t)$ (\ref{eq:inteq}) with initial condition $\varphi_0(x)$ given in (\ref{initial}), using an exponentially convergent algorithm.
The numerical computation is based on expressing the solution $\psi_0(t)$ of (\ref{eq:inteq}) in terms of Chebyshev polynomials.
Since this solution becomes periodic at long times, it is actually convenient to first split time into small intervals, and expand $\psi_0(t)$ into Chebyshev polynomials in each interval.
Then, to compute the right side of (\ref{eq:inteq}), we must compute integrals of $\psi_0(t)$, which we carry out using Gauss quadratures.
Once that is done, (\ref{eq:inteq}) is approximated (by truncating the Chebyshev expansion and the Gauss quadratures) by a finite linear system of equations, which can be solved easily.
One has to pay attention to make sure that this approximation is good.
In this work, we have striven to ensure that the approximation converges to the exact solution exponentially fast as the truncations of the Chebyshev expansion and Gauss-quadratures are removed.
This is not entirely trivial, as both $\psi_0$ and $L$ have square root singularities, so the Chebyshev polynomial expansion and the Gauss quadratures have to be adjusted to take these into account.
\medskip

\indent We take $\frac {\hbar^2k^2}{2m}=\mathcal E_F=4.5$\,eV, $W=5.5$\,eV.
Unless otherwise specified, we also take $\omega=1.55\ \mathrm{eV}$.
The laser period $\tau=\frac{2\pi}\omega$ is then equal to $2.7\ \mathrm{fs}$.
\medskip

\begin{figure}
  \includegraphics[width=8cm]{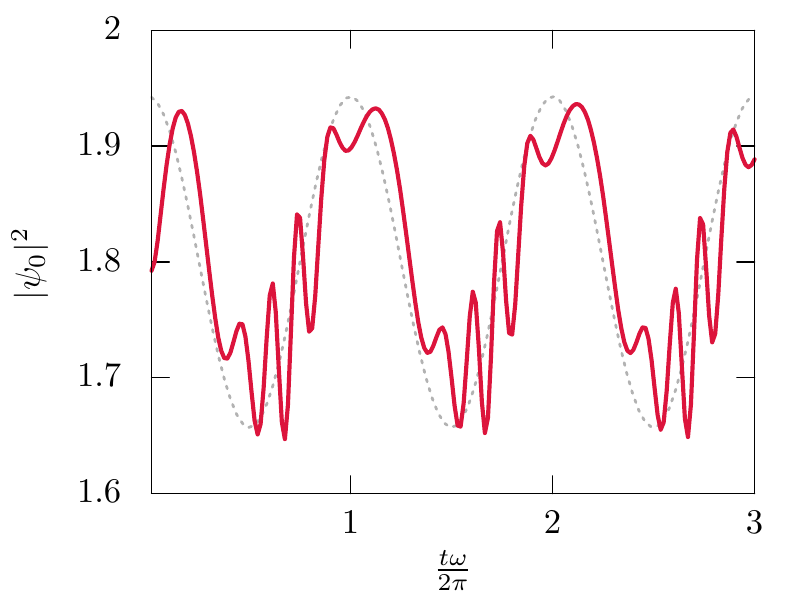}
  \caption{
    The density $|\psi_0|^2$ (recall that $\psi_0(t)\equiv\psi(0,t)$ is the wavefunction at $x=0$) as a function of $\frac{t\omega}{2\pi}$ for 3 periods, with $E=15\ \mathrm V\cdot\mathrm{nm}^{-1}$, $\omega=1.55\ \mathrm{eV}$.
    The dotted line is the graph of $\cos(\omega t)$ (not to scale).
  }
  \label{fig:wavefunction}
\end{figure}

\begin{figure}
  \includegraphics[width=8cm]{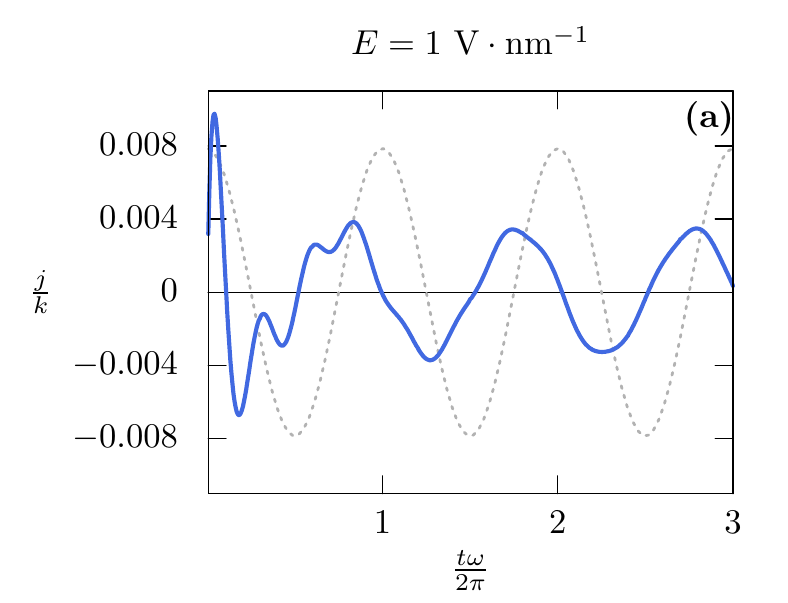}
  \hfill
  \includegraphics[width=8cm]{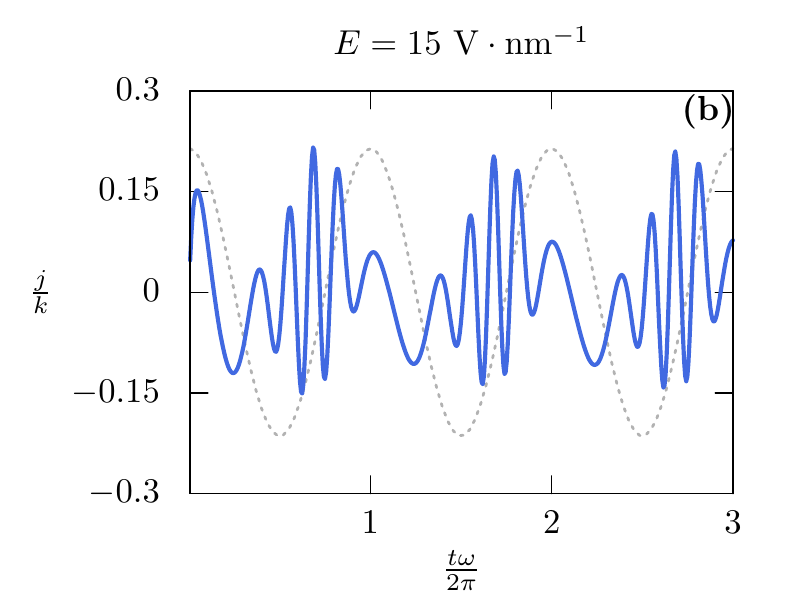}
  \hfill
  \includegraphics[width=8cm]{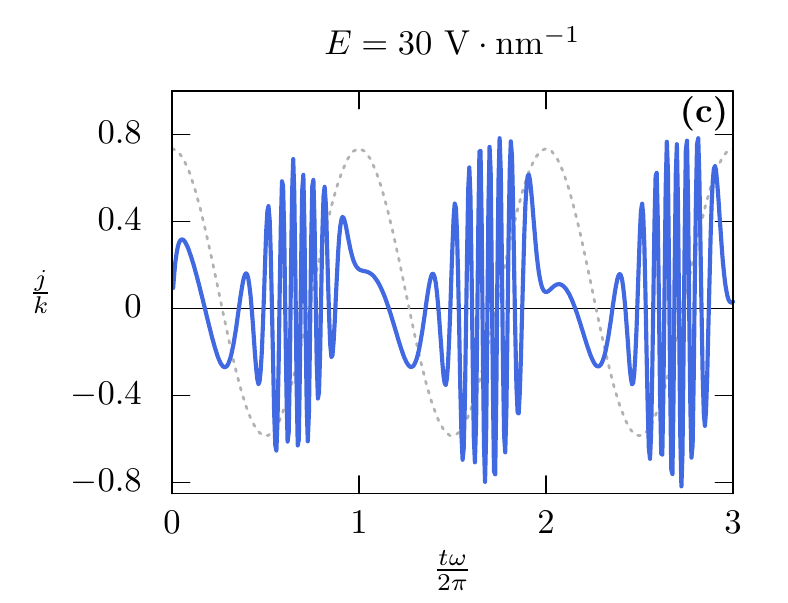}
  \caption{
    The normalized current $\frac jk$ at the interface (recall that we are using atomic units, so $\frac jk$ is dimensionless) as a function of $\frac{t\omega}{2\pi}$ for $\omega=1.55\ \mathrm{eV}$ and three values of the electric field: $E=1,\ 15,\ 30\ \mathrm{V}\cdot\mathrm{nm}^{-1}$.
    The Keldysh parameter $\gamma=\frac{2\omega}E\sqrt W$ for these fields is, respectively, $18.6$, $1.24$ and $0.621$.
    The dotted line is the graph of $\cos(\omega t)$ (not to scale).
    As the field increases, fast oscillations in the current appear.
    These fast oscillations mostly occur while the field is increasing.
  }
  \label{fig:current}
\end{figure}

\begin{figure}
  \includegraphics[width=8cm]{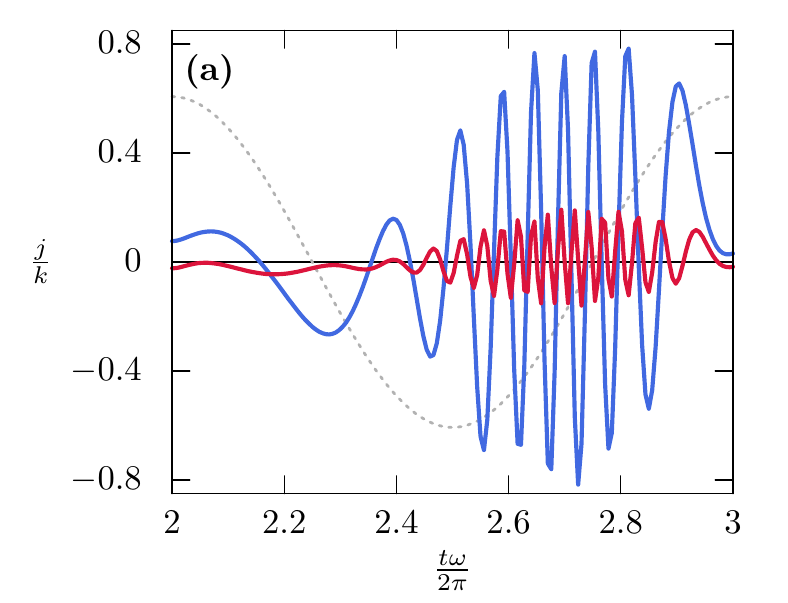}
  \hfill
  \includegraphics[width=8cm]{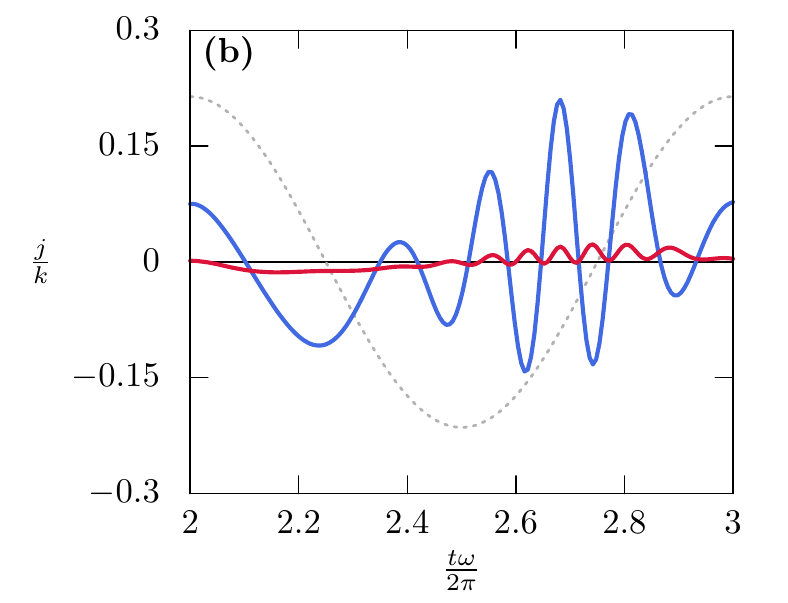}
  \caption{
    The normalized current $\frac jk$ at the interface as a function of $\frac{t\omega}{2\pi}$ for two values of the Keldysh parameter.
    In (a), $\gamma=0.621$, the blue curve has $E=30\ \mathrm{V}\cdot\mathrm{nm}^{-1}$ and $\omega=1.55\ \mathrm{eV}$, and the red curve has $E=15\ \mathrm{V}\cdot\mathrm{nm}^{-1}$ and $\omega=0.755\ \mathrm{eV}$.
    In (b), $\gamma=1.24$, the blue curve has $E=15\ \mathrm{V}\cdot\mathrm{nm}^{-1}$ and $\omega=1.55\ \mathrm{eV}$, and the red curve has $E=7.5\ \mathrm{V}\cdot\mathrm{nm}^{-1}$ and $\omega=0.755\ \mathrm{eV}$.
    At fixed $E$, the frequency of the oscillations decreases with $\omega$ (compare the red curve in (a) with the blue curve in (b)).
    However, this frequency does not only depend on the Keldysh parameter.
  }
  \label{fig:current+}
\end{figure}

\begin{figure}
  \includegraphics[width=8cm]{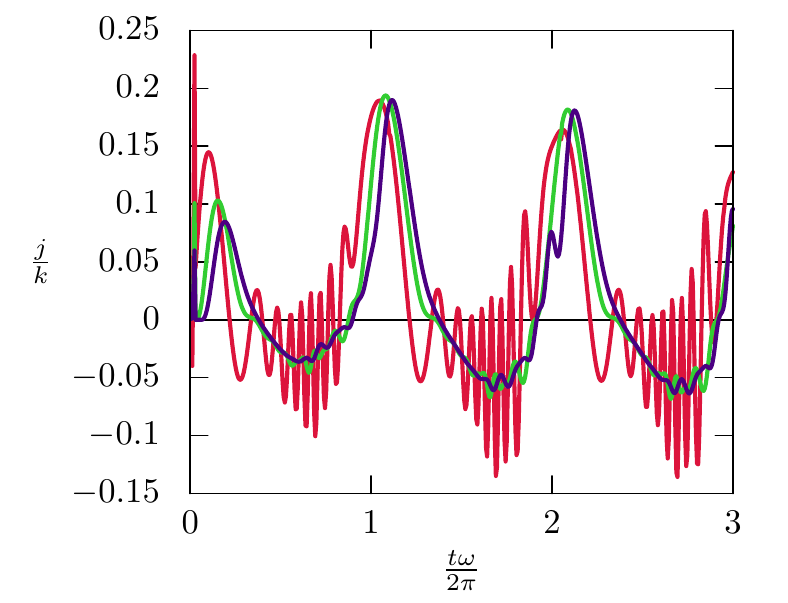}
  \caption{
    The normalized current $\frac jk$ as a function of $\frac{t\omega}{2\pi}$ at positive $x$.
    The parameters here are $E=30\ \mathrm{V}\cdot\mathrm{nm}^{-1}$ and $\omega=1.55\ \mathrm{eV}$, and the values of $x$ are $0.12\ \mathrm{nm}$ (red), $0.24\ \mathrm{nm}$ (green) and $0.37\ \mathrm{nm}$ (purple).
    The fast oscillations die down as $x$ gets larger.
  }
  \label{fig:current-x}
\end{figure}

\indent Figure \ref{fig:wavefunction} shows the density at the interface $|\psi_0(t)|^2$.
The maxima and minima of the density are approximately in phase with the field.
Figure \ref{fig:current} shows the values of the current $j(0,t)$ passing through the origin as a function of time for different strengths of the field, all at $\omega=1.55$\, eV.
We see there a change of behavior as $E$ increases from $E=1$\,V/nm to $E=30$\,V/nm (the Keldysh parameter $\gamma=\frac{2\omega}E\sqrt W$ goes from 18.6 to 0.62).
For large values of $E$, fast oscillations appear, which become faster and larger as $E$ grows.
In Figure \ref{fig:current+}, the current is plotted for various values of $\omega$.
It is seen there that the fast oscillations appear only for small values of $\omega$.
It is also apparent that the frequency of these oscillations is not a function of just the Keldysh parameter.
In Figure \ref{fig:current-x}, we show a plot of the current for positive values of $x$, and see that the fast oscillatory behavior within one period is strongly damped as $x$ increases.
The fact that the electrons cross the surface at different phases of the field does not fit in with the ``simple man'' scenario \cite{KSK92,Co93,KLe18}, where electrons are ejected out of the metal only when the field is positive, or even only when the field is at its maximum.
These oscillations at $x=0$ are also observed in the approximate solution \cite{YGR11}, though the details vary.
We note that the height of the first maximum is linear in $E$, while its location is almost independent of $E$.
Note also that the rapid oscillations occur mostly when the field is increasing.
In the Conclusions we further discuss these oscillations and their possible link to the physics of this process.
\medskip

\begin{figure}
  \includegraphics[width=8cm]{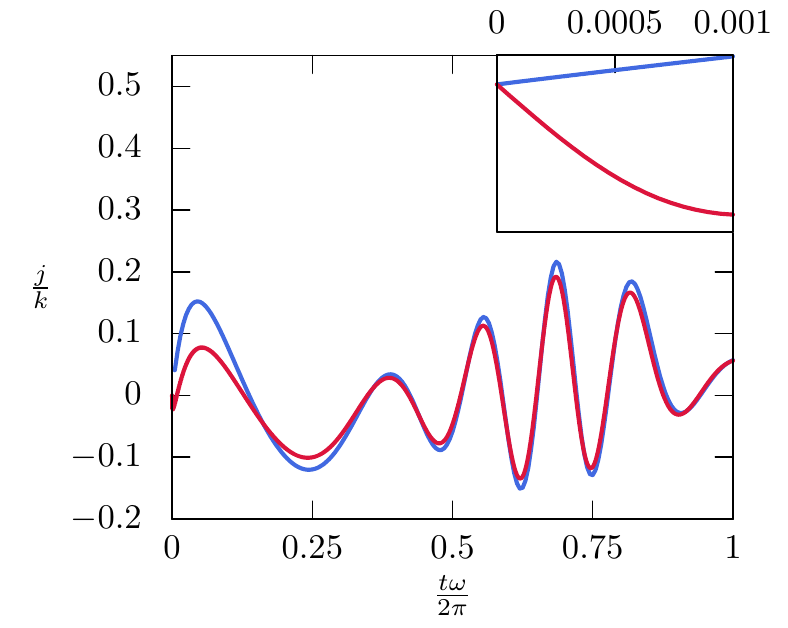}
  \caption{
    The current computed with our method (blue, color online), compared with the Crank-Nicolson algorithm (red), for $\omega=1.55\ \mathrm{eV}$ and $E=15\ \mathrm{V}\cdot\mathrm{nm}^{-1}$.
    The maxima and minima seem to occur at the same time, and the agreement is pretty good for $t>\frac\pi\omega$.
    The inset focuses on short times, for $\frac{t\omega}{2\pi}<0.0005$, for which the Crank-Nicolson algorithm produces a different, and unphysical result: the current initially shoots down to negative values before rising back up.
  }
  \label{fig:cn}
\end{figure}

\indent In Figure \ref{fig:cn}, we show a comparison of our solution with the results obtained from a direct solution of (\ref{schrodinger}) via the Crank-Nicolson algorithm.
The agreement, especially for the location of the maxima and minima after very early times is very good.
It is not so, however, for very short times.
This is expected since the Crank-Nicolson scheme is based on approximating derivatives by finite differences.
However, at short times, $\psi(0,t)\sim t^{\frac32}$, which has a singular second derivative at $0$, so $\partial_t\psi$ is poorly approximated by finite differences.
The fact that the result of the Crank-Nicolson algorithm is different at short times affects the values at later times, as the initial error effectively changes the initial condition.
The agreement at the end of one period is still rather good, indicating that the behavior of the solution behaves weakly on the initial condition.

Note, also, that the Crank-Nicolson method requires a cut-off in $x$, restricting $\psi(x,t)$ to $x\in[-a,a]$.
This causes distortions in $\psi$ due to reflections from these artificial boundaries, and so can only be used for short times (reflections can be avoided by using non-local boundary conditions, such as ``transparent boundary conditions'').
\medskip

\begin{figure}
  \hfil\includegraphics[width=8cm]{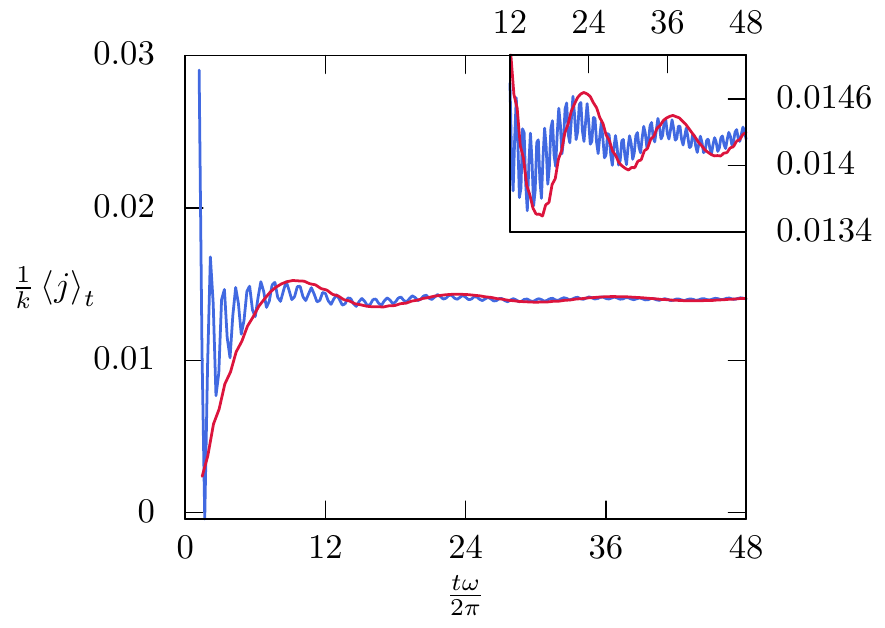}
  \caption{
    The normalized average current $\frac1k\left<j\right>_t$ as a function of $\frac{t\omega}{2\pi}$ at $x=0$ (blue) and $x=0.37\ \mathrm{nm}$ (red) for $\omega=6\ \mathrm{eV}$ and $E=10\ \mathrm{V}\cdot\mathrm{nm}^{-1}$.
    The inset shows the same data, restricted to between the 12th and 48th period.
    Here, $\omega$ is large enough that absorbing one photon suffices to overcome the work function.
    Even after 48 periods the average current has not converged to its asymptotic value.
  }
  \label{fig:avgcurrent}
\end{figure}

\indent In Figure \ref{fig:avgcurrent}, we plot the running average of the current
\begin{equation}
  \left<j\right>_t:=\frac1\tau\int_{t-\tau}^tds\ j(0,s)
  ,\quad
  \tau:=\frac{2\pi}\omega
  .
\end{equation}
We plot $\left<j\right>_t$ at $x=0$ for $\omega=6\ \mathrm{eV}$, $E=10\ \mathrm{V}\cdot\mathrm{nm}^{-1}$, $\gamma=9.6$.
It is seen there that the relative deviation of $\left<j\right>_t$ from its constant asymptotic value, described in section \ref{longtime}, remains significant even when $t\approx48\tau$.
\medskip

\begin{figure}
  \hfil\includegraphics[width=9cm]{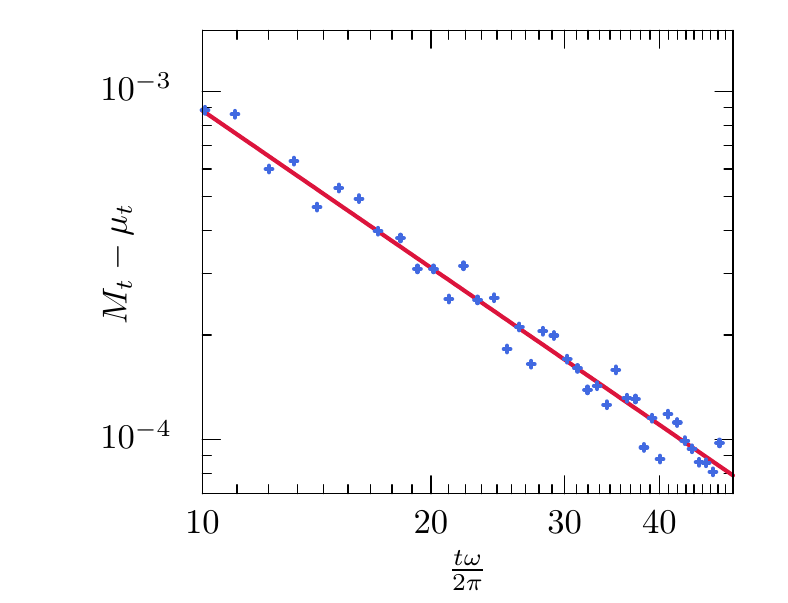}
  \caption{
    The convergence rate to the asymptote of the average current as a function of $\frac{t\omega}{2\pi}$ on a log-log plot for $\omega=6\ \mathrm{eV}$ and $E=10\ \mathrm{V}\cdot\mathrm{nm}^{-1}$.
    The dots are computed at the end of each period $t_n=\frac{2\pi}\omega n$, and their value is the difference between the maximun $M_t$ and the minimum $\mu_t$ of the normalized average current $\frac1k\left<j\right>_t$ in the period immediately preceding $t_n$.
    The red line is a plot of $0.0030\times(\frac{t\omega}{2\pi})^{-\frac32}$, which fits the data rather well.
  }
  \label{fig:decay}
\end{figure}

\indent The rate of the decay of the average current to its asymptotic value is evaluated in Figure \ref{fig:decay}.
In order to compute this rate without having to guess the asymptotic value, we proceed as follows.
At the end of every laser period $t_n=\frac{2\pi}\omega n$, we compute the minimal value $\mu_n$ and maximal value $M_n$ of the average current in the period $(\frac{2\pi}\omega(n-1),\frac{2\pi}\omega n]$.
The plot shows $M_n-\mu_n$ as a function of $t$, and shows that $\left<j\right>_t\approx\left<j\right>_\infty+g(t)t^{-\frac32}$, where $g(t)$ is bounded and asymptotically constant.
This is consistent with the exact result in section \ref{longtime}.
\medskip

\begin{figure}
  \hfil\includegraphics[width=8cm]{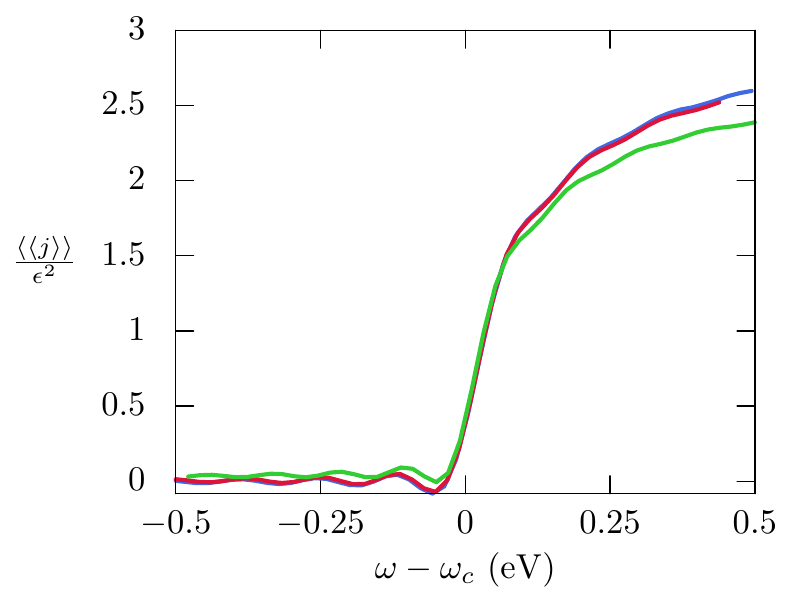}
  \caption{
    The average of the average of the current $\frac1{\epsilon^2}\left<\left<j\right>_t\right>$ after 12 periods as a function of $\omega-\omega_c$, for various values of the field: $E=3\ \mathrm{V}\cdot\mathrm{nm}^{-1}$ (blue), $E=10\ \mathrm{V}\cdot\mathrm{nm}^{-1}$ (red),  $E=30\ \mathrm{V}\cdot\mathrm{nm}^{-1}$ (green).
    We see a sharp transition as $\omega$ crosses $\omega_c$.
  }
  \label{fig:omega}
\end{figure}

\indent In Figure \ref{fig:omega}, we show an estimate of the asymptotic average current as a function of $\omega$ in the vicinity of the one-photon threshold $\omega_c=W+\frac{E^2}{4\omega^2}$ for $E=3,10,30\ \mathrm{V}\cdot\mathrm{nm}^{-1}$.
In order to reduce the fluctuations and estimate the long-time average current, we took a second average, and computed the average over a laser period of the average current, defined as
\begin{equation}
  \left<\left<j\right>\right>:=\frac1\tau\int_{T-\tau}^T dt\ \left<j\right>_t
  .
\end{equation}
By dividing the current by $\epsilon^2$, we see that the the average of the average of the current is proportional to $\epsilon^2$.
We see that there is a steep increase in $\left<\left<j\right>\right>$ as $\omega$ increases past $\omega_c$.
This is precisely what is observed in experiments on the photoelectric effect, where the emission of electrons from the metal surface has such a threshold \cite{Mi16}.
This became a key element in Einstein's ansatz of localized photons.
Here, in the classical treatment of the laser field, this phenomenon appears as a consequence of the quantum treatment of the electrons.
It shows that, despite it simplicity, this model captures essential features of the physical phenomena.
\medskip

\section{Long time behavior of $\psi(x,t)$}\label{longtime}
\indent The long time behavior of the system is given by the poles of the Laplace transform
\begin{equation}
  \hat{\psi}(x,p)=\int_0^{\infty}dt\ e^{-pt} \psi(x,t)
\end{equation}
on the imaginary axis, as can be seen by taking the inverse Laplace transform
\begin{equation}
  \psi(x,t)=\int_{-i\infty}^{i\infty}dp\ e^{pt}\hat\psi(x,p)
\end{equation}
where the path of integration can be deformed to get contributions only from poles and branch cuts in the negative real half-plane $\mathcal Re(p)\leqslant 0$.
The poles that have negative real parts would give rise to exponentially decaying terms while branch cuts generally contribute $t^{-\frac32}$ terms to the approach to the asymptotic state.
The contribution from poles on the imaginary $p$-axis then give the long-time asymptotics of $\psi$ is of the form 
\begin{equation}\label{4p1}
\psi(x,t)\sim {\rm e}^{\frac 12 i k^2 t} \bar{\psi}(x,t)
\end{equation}
where $\bar{\psi}(x,t)$ is periodic in $t$ with period $2\pi/\omega$ and $\frac 12  k^2$ is the energy of the electrons in the incoming beam.
This corresponds to the poles of $\hat{\psi}(p,t)$ on the imaginary axis being at 
\begin{equation}\label{4p2}
p=-\frac 12  i k^2+i\omega n,\ \ \ n\mathrm{\ \ integer}
\end{equation}
\medskip

\indent As shown in \cite{CCe}, $\bar{\psi}(x,t)$ coincides with the wave function $\mathcal{U}(x,t)$ computed by Faisal et al \cite{FKS05}.
In that work it was assumed that (\ref{schrodinger}) has a solution of the form (\ref{4p1}) with an incoming beam $e^{ikx}$ for $x<0$, without considering any initial conditions.
Computing the residues at the poles on the imaginary axis we show that, for $x<0$,
\begin{equation}
  \bar\psi(x,t)\sim
  e^{ikx}
  +\sum_{m\in\mathbb Z}e^{-im\omega t}e^{-ix\sqrt{k^2+2m\omega}}\mathcal R_m
  \label{asym1}
\end{equation}
and for $x>0$,
\begin{equation}
  \bar\psi(x,t)\sim
  e^{i\frac E\omega x\sin\omega t}\sum_{n,m\in\mathbb Z}e^{-in\omega t}g_{n-m}^{(\kappa_m)}e^{-\kappa_mx}\mathcal T_m
  \label{asym2}
\end{equation}
where
\begin{equation}
  \kappa_m=\sqrt{2U-k^2+\frac{E^2}{2\omega^2}-2m\omega}
  \label{kappa}
\end{equation}
and
\begin{equation}
  g_{n-m}^{(\kappa_m)}=\frac\omega{2\pi}\int_0^{\frac{2\pi}\omega}dt\ e^{-i(n-m)\omega t}e^{\frac{i\frac{E^2}{4\omega^2}}\omega\sin(2\omega t)+\kappa_m\frac{2E}{\omega^2}\cos(\omega t)}
  .
\end{equation}
This is exactly of the form obtained in \cite{FKS05}.
$\mathcal R_n$ and $\mathcal T_m$ are computed by matching boundary values of $\psi(x,t)$ and $\partial_x\psi(x,t)$ at $x=0$.
The phase $e^{i\frac E\omega x\sin\omega t}$ comes from a change of gauge with respect to \cite{FKS05} (we use the ``length'' gauge, instead of the ``magnetic gauge'' \cite{CFe87}).
\medskip

\indent A physical interpretation of (\ref{asym1})-(\ref{asym2}), see \cite{FKS05}, is that an electron in a beam coming from $-\infty$ and moving in the positive $x$-direction, ${\rm e}^{ikx},\ k>0$, absorbs or emits ``$m$ photons'' and is either reflected, transmitted or damped.
Transmission occurs when $m\omega>U+\frac {E^2}{4\omega^2}-\frac{k^2}2\equiv\omega_c$.
Since we are taking $\frac{k^2}2=\mathcal E_F$, we have that $U-\frac{k^2}2=W$, the work function.
The $\frac{E^2}{4\omega^2}$ in (\ref{kappa}) term corresponds to the ponderomotive energy of the electron \cite{Wo35} in the laser field.
Damping occurs when $m\omega<\omega_c$.
$\omega_c$ is the minimum frequency necessary to let the electron with incoming kinetic energy $\frac12k^2$ (in the $x$-direction) propagate to the right of the potential barrier.
For large $x>0$, the current in the $m$-photon channel will have kinetic energy $m\omega-\omega_c$ and the current will be given by $\sqrt{m\omega-\omega_c}$.
This will also be equal to the average current at large $t$, which is independent of $x$.
This explains the picture in Figure \ref{fig:omega} for $m=1$.
The larger $m$ values necessary for smaller $\omega$ are difficult to see.
\medskip

\section{Concluding remarks}
\indent In this paper we presented, for the first time, the exact solution of the time-dependent Schr\"odinger equation (\ref{schrodinger}).
This is the simplest physical model describing the emission of electrons from a flat metal surface by an oscillating field.
The model was first used by Fowler and Nordheim \cite{FN28} for emission by a constant electric field, $\omega=0$.
Their formula for the steady state current, obtained from the stationary solution of (\ref{schrodinger}), with $\omega=0$, still forms the basis of the interpretation of experiments at present time \cite{Je17}.
There are modifications due to the Shottky effect, but these are not expected to change the basic results.
The situation is different when the field acting on the metal surface is periodic in time.
Equation (\ref{schrodinger}) no longer has an explicit ``stationary'' (in this case, periodic) solution of the type considered by Faisal et al \cite{FKS05}.
In fact, even the existence of physical solutions of (\ref{schrodinger}), i.e. ones bounded for all $x$, is problematic from a mathematical point of view.
This is what we establish here by proving that the integral equation (\ref{eq:inteq}) indeed has solutions which give a physical $\psi(x,t)$ for all $t>0$.
We prove this for a very general class of initial conditions and carry out exact numerical solutions for the case of a particular physically motivated initial state.
The numerics are proven to give arbitrary accuracy for any fixed $t$ and specified bounded initial state.
\medskip

\indent Our results reveal, as shown in the figures, many new features of the exact solution of (\ref{schrodinger}), e.g. the slow convergence in time of the average of the current to its asymptotic value, and the rapid oscillations at the interface for strong fields and small $\omega$.

A more detailed examination of our solution shows that the rapid oscillations are confined to a very narrow region close to the metal surface. A time Fourier transform of the wave function -- which corresponds to looking in energy space -- indicates that these fast oscillations are due to energy absorptions,  $E_n=n \hbar\omega$ for all  $n$ such that $E_n$ exceeds the work function plus the ponderomotive energy.  Farther away from the metal surface, due to the transition to a semiclassical behavior, energy absorption  and hence the rapid oscillations, cease rapidly. The purely quantum processes occur in the tunneling region proximal to  the surface.

The slow convergence of the average of the current indicates that different initial conditions may give different results in short pulse experiments.
On the other hand, we prove that there is indeed an asymptotic periodic state of the form assumed by Faisal et al. \cite{FKS05}.
The asymptotic form (\ref{asym1})-(\ref{asym2}) is true for all initial conditions of the form $\Theta(-x)e^{ikx}+f_0(x)$ as long as $f_0(x)$ only contains terms which are square integrable.
The additional terms in $\psi(x,t)$ which come from $f_0(x)$ go to zero as $t\to\infty$.
This follows from the fact, proven in \cite{CCe}, that the Floquet operator associated to (\ref{schrodinger}) has no point spectrum.
\medskip

\begin{acknowledgements}
We thank David Huse, Kevin Jensen and Donald Schiffler for very valuable discussions, as well as the anonymous referees who made invaluable comments and helped improve this paper. This work was supported by AFOSR Grant No. FA9500-16-1-0037. O.C. was partially supported by the NSF-DMS (Grant No. 1515755). I.J. was partially supported by the NSF-DMS (Grants No. 31128155 and 1802170). J.L.L. thanks the Institute for Advanced Study for its hospitality.    
\end{acknowledgements}

\bibliographystyle{apsrev}
\bibliography{bibliography}

\appendix
\section{The exact solution of \eqref{schrodinger}}

Let $\psi_0(t)=\psi(0,t)$ and $\partial_x\psi_{0}(t)=\partial_x\psi(x,t)|_{x=0}$. 
 The operator $L$ in (\ref{eq:inteq}) is given by
\begin{equation}
  \begin{array}{>\displaystyle l}
    L\psi_0(t):=
    \frac{E}{2\omega\sqrt{2i\pi}}\int_0^tds\ \psi_0(s)\frac{\alpha(s,t)}{\sqrt{t-s}}e^{if(s,t)}
    \\[0.5cm]\hfill
    +\frac1{2\pi}\int_0^t du\ \psi_0(u)\int_u^t ds\ \frac{g(s,t)}{\sqrt{s-u}}
  \end{array}
\end{equation}
where 
\begin{equation}
  \alpha(s,t):=\sin(\omega s)+\frac{\cos(\omega t)-\cos(\omega s)}{\omega(t-s)}
  \label{alpha}
\end{equation}
\begin{equation}
  \begin{array}{>\displaystyle l}
    f(s,t):=
    \frac{E^2(\cos(\omega t)-\cos(\omega s))^2}{2\omega^4(t-s)}
    \\[0.5cm]\hfill
    -\left(V+\frac{E^2}{4\omega^2}\right)(t-s)
    +\frac{E^2}{8\omega^3}(\sin(2\omega t)-\sin(2\omega s))
  \end{array}
\end{equation}
and
\begin{equation}
  g(s,t):=
  \frac{e^{if(s,t)}-1}{2(t-s)^{\frac32}}+\frac{i\partial_sf(s,t)e^{if(s,t)}}{\sqrt{t-s}}
  .
\end{equation}

The function  $h$ in (\ref{eq:inteq}) is given by
\begin{equation}
  \begin{array}{r@{\ }>\displaystyle l}
    h(t):=&h_+(0,t)+h_-(0,t)
    \\[0.5cm]&
    -\frac1{\pi}\int_0^tdu\ h_-(u)\int_u^t ds\ \frac{g(s,t)}{\sqrt{s-u}}
  \end{array}
\end{equation}
where
\begin{equation}
  h_-(x,t):=\frac{e^{-\frac{i\pi}4}}{\sqrt{2\pi  t}}\int_{-\infty}^0 dy\ \varphi_0(y)e^{\frac i{2t}(x-y)^2}
\end{equation}
and
\begin{equation}
  \begin{array}{r@{\ }>\displaystyle l}
    h_+(x,t):=&
    \frac{e^{-\frac{i\pi}4}}{\sqrt{2\pi  t}}
    e^{i\frac E\omega x\sin(\omega t)-i(V+\frac{E^2}{4\omega^2})t+\frac{iE^2}{8\omega^3}\sin(2\omega t)}
    \\[0.5cm]&\times
    \int_0^\infty dy\ \varphi_0(y)
    e^{\frac i{2t}(-x+y+\frac{E}{\omega^2}(1-\cos(\omega t)))^2}
  \end{array}
\end{equation}
For our choice (\ref{initial}) of $\varphi_0$, these two functions are explicit:
\begin{equation}
  \begin{array}{>\displaystyle l}
    h_-(x,t)=\frac{e^{-\frac{ik^2}{2m}t}}2\Big(
      e^{ikx}
      \mathrm{erfc}({\textstyle e^{-\frac{i\pi}4}(-\sqrt{\frac t2}k+\frac1{\sqrt{2t}}x})
      \\[0.5cm]\hfill
      +\frac{ik+\sqrt{2U-k^2}}{ik-\sqrt{2U-k^2}}
      e^{-ikx}
      \mathrm{erfc}({\textstyle e^{-\frac{i\pi}4}(\sqrt{\frac t2}k+\frac1{\sqrt{2t}}x)})
    \Big)
    .
  \end{array}
\end{equation}
\medskip
and
\begin{equation}
  \begin{array}{>\displaystyle l}
    h_+(x,t)=
    \frac{ik}{ik-\sqrt{2U-k^2}}
    e^{i\frac E\omega\sin(\omega t)x-\sqrt{2U-k^2}x}
    \\[0.5cm]\hfill\times
    e^{\frac{E}{\omega^2}(1-\cos(\omega t))\sqrt{2U-k^2}-i(\frac{k^2}2+\frac{E^2}{4\omega^2})t+i\frac{E^2}{8\omega^3}\sin(2\omega t)}
    \\[0.5cm]\hfill\times
    \mathrm{erfc}(e^{-\frac{i\pi}4}({\textstyle i\sqrt{\frac t2}\sqrt{2U-k^2}+\frac E{\omega^2}\frac{1-\cos(\omega t)}{\sqrt{2t}}-\frac1{\sqrt{2t}}x}))
    .
  \end{array}
\end{equation}
$\hat{\psi}_-$ is obtained by explicitly solving (\ref{A}):
\begin{equation}\begin{array}{c}
\hat{\psi}_-(\xi,t)= e^{-i\frac{\xi^2}2t} \frac{1}{\sqrt{2\pi}}\int_{-\infty}^0 e^{-i\xi x}\varphi_0(x)\, dx
\\ \\ 
+\frac1{2\sqrt{2\pi}}  \int_0^t e^{-i\frac{\xi^2}2(t-s)}\ \left[  i\partial_x\psi_0(s)-\xi \psi_0(s) \right]\, ds
\end{array}\end{equation}
The PDE  (\ref{B}) can be solved explicitly by characteristics to give  $\hat{\psi}_+$:
 \begin{equation}
 \hat{\psi}_+(\xi,t) =G(\xi- \frac E{\omega}\sin \omega t,t)
 \end{equation}
  where
  \begin{multline}
  G(u,t)=e^{-i\Phi(u,t)} \frac{1}{\sqrt{2\pi}}\int_0^{\infty}e^{-ikx}\varphi_0(x)\, dx +
  \\
  +\frac1{2\sqrt{2\pi}}\int_0^t ds\, e^{-i(\Phi(u,t)-\Phi(u,s))}\cdot\hfill\\
  \cdot\left[ -i \partial_x\psi_0(s)+(u+\frac E{\omega}\sin (\omega s)) \psi_0(s)\right]
  \end{multline}
where
\begin{multline}
\Phi(u,t)=\\
=\left(\frac{u^2}2  +V+\frac{E^2}{4\omega^2}\right)t + u \frac{E}{\omega^2}(1-\cos(\omega t))-\frac{E^2}{8\omega^3} \sin(2\omega t)
.
\end{multline}
One can then check that
\begin{equation}
  \label{eq:postconvo}
    \partial_x\psi_0(t)=\frac{\sqrt2}{\sqrt{i \pi}} \frac d{dt}\left[ \psi_0(t)*t^{-1/2}- 2 h_-(0,t)*t^{-1/2}\right]   
 \end{equation}
where '*' denotes the Laplace convolution
\begin{equation}
  [f*g](t)=\int_0^t ds\ f(s)g(t-s)
\end{equation}
is continuous for $t>0$.
\bigskip 
  
The solution $\psi(x,t)$ of the Schr\"odinger equation (\ref{schrodinger}) is, for $x<0$, the inverse Fourier transform of $\hat{\psi}_-$, while for $x>0$ it equals the inverse Fourier transform of $\hat{\psi}_+$. Namely,
\begin{multline}
  \label{eq:psimxt}
  {\psi}_-(x,t)=h_-(x,t)+\\+ \frac{e^{\frac{i\pi}4}}{2\sqrt{2\pi}} \int_0^t ds\, \left(\partial_x\psi_0(s)+i\psi_0(s)\frac x{t-s}\right)\frac{e^{\frac{ix^2}{2(t-s)}}}{\sqrt{t-s}}\\
\end{multline}
and
\begin{multline}
  \psi_+(x,t)=h_+(x,t)-\\
  -e^{i\frac E\omega\sin(\omega t)x}\frac{e^{\frac{3i\pi}4}}{2\sqrt{2\pi}}\int_0^t ds\frac{\Gamma_+(s,x,t)}{\sqrt{t-s}}e^{iF(x,s,t)}
\end{multline}
where
\begin{multline}
  \label{eq:h0}
  \Gamma_+(s,x,t):=
  -i\partial_x\psi_0(s)+\\
  +\left(\frac E\omega\sin(\omega s)+\frac E{\omega^2}\frac{\cos(\omega t)-\cos(\omega s)}{t-s}+\frac x{t-s}\right)\psi_0(s)
\end{multline}
and
\begin{equation}
  \label{eq:defx}
  F(x,s,t)=f(s,t)+x\frac{E}{\omega^2}\frac{\cos(\omega t)-\cos(\omega s)}{t-s}
  +\frac{ix^2}{2(t-s)}
  .
\end{equation}
\end{document}